\documentclass[10pt]{iopart}
\expandafter\let\csname equation*\endcsname\relax
\expandafter\let\csname endequation*\endcsname\relax

\usepackage {amsmath}
\usepackage{graphicx}
\usepackage{lipsum}
\usepackage{bm}
\usepackage{tabu}
\usepackage[utf8]{inputenc}
\usepackage[T1]{fontenc}
\usepackage{mathptmx}
\usepackage{color}
\usepackage{xcolor}
\usepackage{cite}
\usepackage{iopams}  

\newcommand{\Me}{\text{M}^e}
\newcommand{\Mo}{\text{M}^o}
\newcommand{\MGe}{\text{MG}^e}
\newcommand{\MGo}{\text{MG}^o}
\newcommand{\HMGp}{\text{HMG}^+}
\newcommand{\HMGm}{\text{HMG}^-}
\newcommand{\HMGV}{{\bf HMGV}}

\newcommand{\GB}{\text{GB}}

\newcommand{\Je}{\text{Je}}
\newcommand{\Jo}{\text{Jo}}
\newcommand{\ce}{\text{ce}}
\newcommand{\se}{\text{se}}

\begin{document}
\title[Experimental generation of Helical Mathieu-Gauss vector modes]{Experimental generation of Helical Mathieu-Gauss vector modes}

\author{Carmelo Rosales-Guzm\'an$^{1,2}$, Xiao-Bo Hu$^1$, Valeria Rodr\'iguez-Fajardo$^3$, Raul I. Hernandez-Aranda$^4$, Andrew Forbes$^3$ and Benjamin Perez-Garcia$^4$}
\address{$^1$ Wang Da-Heng Collaborative Innovation Center, Heilongjiang Provincial Key Laboratory of Quantum Manipulation and Control, Harbin University of Science and Technology, Harbin 150080, China}
\address{$^2$ Centro de Investigaciones en Óptica, A.C., Loma del Bosque 115, Colonia Lomas del campestre, C.P. 37150 León, Guanajuato, Mexico}
\address{$^3$ School of Physics, University of the Witwatersrand, Private Bag 3, Johannesburg 2050, South Africa}
\address{$^4$ Photonics and Mathematical Optics Group, Tecnologico de Monterrey, Monterrey Mexico}
\ead{carmelorosalesg@hrbust.edu.cn}

\begin{abstract}
Vector modes represent the most general state of light in which, the spatial and polarisation degrees of freedom are coupled in a non-separable way. Crucially, while polarisation is limited to a bi-dimensional space, the spatial degree of freedom can take any spatial profile. However, most generation and application techniques are mainly limited to spatial modes with polar cylindrical symmetry, such as Laguerre- and Bessel-Gauss modes. In this manuscript we put forward a novel class of vector modes with its spatial degree of freedom encoded in the set of helical Mathieu-Gauss beams of the elliptical cylindrical coordinates. We first introduce these modes theoretically and outline their geometric representation on the higher-order Poincar\'e sphere. Later on, we demonstrate their experimental generation using a polarisation-insensitive technique comprising the use of a digital micromirror device. Finally, we provide with a qualitative and a quantitative characterisation of the same using modern approaches based on quantum mechanics tools. It is worth mentioning that non-polar vector beams are highly desired in various applications, such as optical trapping and optical communications. 
\end{abstract}

\ioptwocol
\maketitle
\section{Introduction}
Complex vector modes have gained popularity in recent time, partly fuelled by the plethora of applications they span \cite{Rosales2018Review,Roadmap}, but also due to the many similarities they hold with quantum entangled states \cite{Spreeuw1998,Qian2011,Aiello2015,konrad2019,forbes2019classically}. This similarities originate from the non-separable coupling between the spatial and polarisation degrees of freedom \cite{Otte2016,Galvez2015,Bauer2015,Beckley2010,Otte2018}, and as a result the transverse polarisation distributions of vector modes become non-homogeneous, thus giving rise to a wide variety of exotic polarisation patterns. Crucially, even though polarisation is restricted to a two-dimensional space, the spatial shape has theoretically infinite possibilities. In spite of that, the vast majority of studies have only considered modes with polar cylindrical symmetry, such as,  Bessel and Laguerre-Gaussian modes \cite{Zhan2009,Dudley2013,Otte2018b,Otte2018}. It is therefore not surprising that most applications revolve around cylindrical vector modes \cite{Bhebhe2018,Hu2019,BergJohansen2015,Li2016,Ndagano2017,Wang2015}. In a similar way many generation and characterisation techniques apply only to cylindrical vector modes \cite{Ndagano2016,Zhaobo2019,Ndagano2018,Ndagano2017ModeSorter,MorenoSR2017}. Notably, recent technological advances have put at our disposal computer-controlled devices capable to generate light beams with arbitrary spatial shapes. Example of such are Spatial Light Modulators (SLMs) \cite{SPIEbook,Forbes2016,Clark2016,Maurer2007} and Digital Micromirror Devices (DMDs) \cite{Scholes2019,Ren2015,Rosales2020,Mitchell2016,Mirhosseini2013,Xu2014,Gong2014,Gong2013,Yao-Li2020}, which have gained popularity in the last two decades. It should be possible then to generate vector beams with non-polar symmetry. Nonetheless, only few examples exist so far, Airy-vortex \cite{Zhou2015} and Ince-Gaussian \cite{Otte2018a, Yao-Li2020} vector modes amongst them. 

Here, we introduce a novel class of vector modes with an elliptical spatial distribution given in terms of the helical Mathieu beams \cite{Gutierrez-Vega2000,Gutierrez-vega2001,Gutierrez-Vega2003,Chavez-Cerda2002,Rodriguez-Lara2008,Hernandez2010}, which we term Helical Mathieu-Gauss vector ({\bf HMGV}) modes. Even though these modes are theoretically well-known, since they are solutions of the paraxial wave equation in its vector form \cite{Chafiq2007}, their experimental generation has been only briefly discussed in literature \cite{Gong2014}. While scalar beams with elliptical symmetry have already demonstrated their potential in various applications \cite{Lopez-Mariscal2006,Gu2020,Sakpal2018,Otte2019}, the {\bf HMGV} modes introduced here incorporate another degree of freedom, paving the way for novel applications.

The manuscript is organised as follows. In  Section \ref{Theory} we present a mathematical description of {\bf HMGV} modes starting with a brief description of the scalar Mathieu-Gauss beams and ending with their geometric representation on the higher-order Poincar\'e sphere. In section \ref{Experiment} we delve into the experimental generation of such modes, which consist of a non-separable weighted superposition of both degrees of freedom, implemented with a Digital Micromirror Device (DMD) \cite{Rosales2020}. Here we also introduce the techniques we utilise in the last section to characterise the generated modes, Stokes polarimetry \cite{Zhaobo2019} and concurrence \cite{Selyem2019}. Finally in section \ref{results} we present a detailed analysis of a representative set of {\bf HMGV} modes where we explore some of the parameters that define the aforementioned modes. 

\section{Theory}\label{Theory}

\subsection{Scalar Helical Mathieu-Gauss beams}
Mathieu beams are found as solutions of the Helmholtz equation in elliptical cylindrical coordinates $(\xi,\eta,z)$, where it separates into a longitudinal and a transverse part \cite{Gutierrez-Vega2000}. The former having a solution of the form $\exp(-ik_zz)$ while the latter being a solution of the equation
\begin{equation}
   \left[\frac{\partial^2}{\partial\xi^2}+\frac{\partial^2}{\partial\eta^2}+\frac{f^2k_t^2}{2}(\cosh2\xi-\cos2\eta)\right]u_T(\xi,\eta)=0.
   \label{Eq:Mathieu}
\end{equation}
Here $\xi \in[0,\infty)$ plays the analogous role of the radial coordinate, $\eta\in[0,2\pi)$ of the angular coordinate and $z\in(-\infty,\infty)$ is the propagation coordinate. The parameter $f$ is the semi-focal distance defined in terms of the  major and  minor axis, $a$  and $b$ respectively, as $f^2=a^2-b^2$ and is related to the eccentricity $e$ by $e=f/a$. Moreover, the Cartesian coordinates $(x,y,z)$ relate to the elliptical coordinates $(\xi,\eta, z)$ as $x=f\cosh\xi\cos\eta$, $y=\sinh\xi\sin\eta$ and $z=z$. The parameters $k_z$ and $k_t$ are the longitudinal and transverse components of the wave vector $\bf k$, whose magnitude $k=2\pi/\lambda$ satisfies the relation $k^2=k_t^2+k_z^2$. Equation \ref{Eq:Mathieu} can be solved using separation of variables, allowing  to split this equation into the radial and the angular Mathieu equations \cite{NIST2010}
\begin{equation}
\begin{split}
    \left[\frac{d^2}{d\xi^2}-(a-2q\cosh2\xi)\right]R(\xi) &= 0\\
    \left[\frac{d^2}{d\eta^2}-(a-2q\cos2\eta)\right]\Theta(\eta) &= 0,
\end{split}
\end{equation}
where $q$ is a dimensionless parameter related to the transverse component of the wave vector $k_t$ as $q=(fk_t/2)^2$.  The solutions to this equation are the radial and angular Mathieu functions respectively, which are given by
\begin{equation}\label{MGscalar}
\begin{split}
    \Me_m(\xi,\eta;q) &= C_m \Je_m(\xi,q)\ce_m(\eta,q),\\
    \Mo_m(\xi,\eta;q) &= S_m \Jo_m(\xi,q)\se_m(\eta,q),\\
\end{split}
\end{equation}
where $C_m$, $S_m$ are normalisation constants. The $m$th-order functions $\Je_m$ and $\Jo_m$, are the even and odd radial Mathieu Functions, while the functions $\ce_m$ and $\se_m$ are the even and odd angular Mathieu functions. The sub-index $m$ indicates the order of the function and is a non-negative integer, $m=0, 1, 2, 3, ...$, for even modes and a positive integer for odd modes, $m=1, 2, 3, ...$. The modes described by Eq.~\ref{MGscalar} are the non-diffracting Mathieu beams, which carry an infinite amount of energy and therefore cannot be realised experimentally. A finite-energy version of such modes are the Mathieu-Gauss (MG) beams, which retain the non-diffracting properties of the ideal Mathieu beams over a finite propagation distance $[-z_{max},z_{max}]$, where $z_{max}=\omega_0k/k_t$. For $|z|\geq z_{max}$ the diffraction properties of the Gaussian envelope becomes dominant and the MG modes diverge, acquiring a ring-shaped far-field intensity distribution. Mathematically, the MG beams are given by \cite{Gutierrez-Vega2005},
\begin{equation}\label{MGmodes}
    \begin{split}
    \MGe_{m}(\tilde{\xi},\tilde{\eta},z;q) &=\exp\left({-\frac{ik_t^2}{2k}\frac{z}{\mu}}\right)\GB({\bf r})\Me_{m}(\tilde{\xi},\tilde{\eta};q),\\
    \MGo_{m}(\tilde{\xi},\tilde{\eta},z;q) &=\exp\left({-\frac{ik_t^2}{2k}\frac{z}{\mu}}\right)\GB({\bf r})\Mo_{m}(\tilde{\xi},\tilde{\eta};q),\\
    \end{split}
\end{equation}
where the Cartesian coordinates $(x,y)$ are redefined in terms of the  complex elliptic variables $(\tilde{\xi},\tilde{\eta})$ as, $x=f_0(1+iz/z_R)\cosh\tilde{\xi}\cos\tilde{\eta}$ and $y=f_0(1+iz/z_R)$ $\sinh\tilde{\xi}\sin\tilde{\eta}$ with $f_0$ the semifocal separation at $z=0$. Further, the term $\GB({\bf r})$ is the fundamental Gaussian beam defined as,
\begin{equation}\label{MGmoes}
    \GB({\bf r})=\exp\left({-\frac{r^2}{\mu\omega_0^2}}\right)\frac{\exp(ikz)}{\mu}.\\
\end{equation}
The parameter $\mu=\mu(z)$ is defined as $\mu=1+iz/z_R$, with $z_R=k\omega_0^2/2$ being the Rayleigh range of a Gaussian beam with waist radius $\omega_0$.

Suitable combinations of the modes defined above gives rise to the Helical Mathieu-Gauss (HMG) modes, which can be written mathematically as
\begin{equation}\label{helical}
\begin{split}
    \HMGp_m(\tilde{\xi},\tilde{\eta},z;q) = \MGe_m(\tilde{\xi},\tilde{\eta},z;q) + i \MGo_m(\tilde{\xi},\tilde{\eta},z;q),\\
    \HMGm_m(\tilde{\xi},\tilde{\eta},z;q) = \MGe_m(\tilde{\xi},\tilde{\eta},z;q) - i \MGo_m(\tilde{\xi},\tilde{\eta},z;q).
 \end{split}
\end{equation}

\begin{figure}[b]
    \centering
    \includegraphics[width=0.49\textwidth]{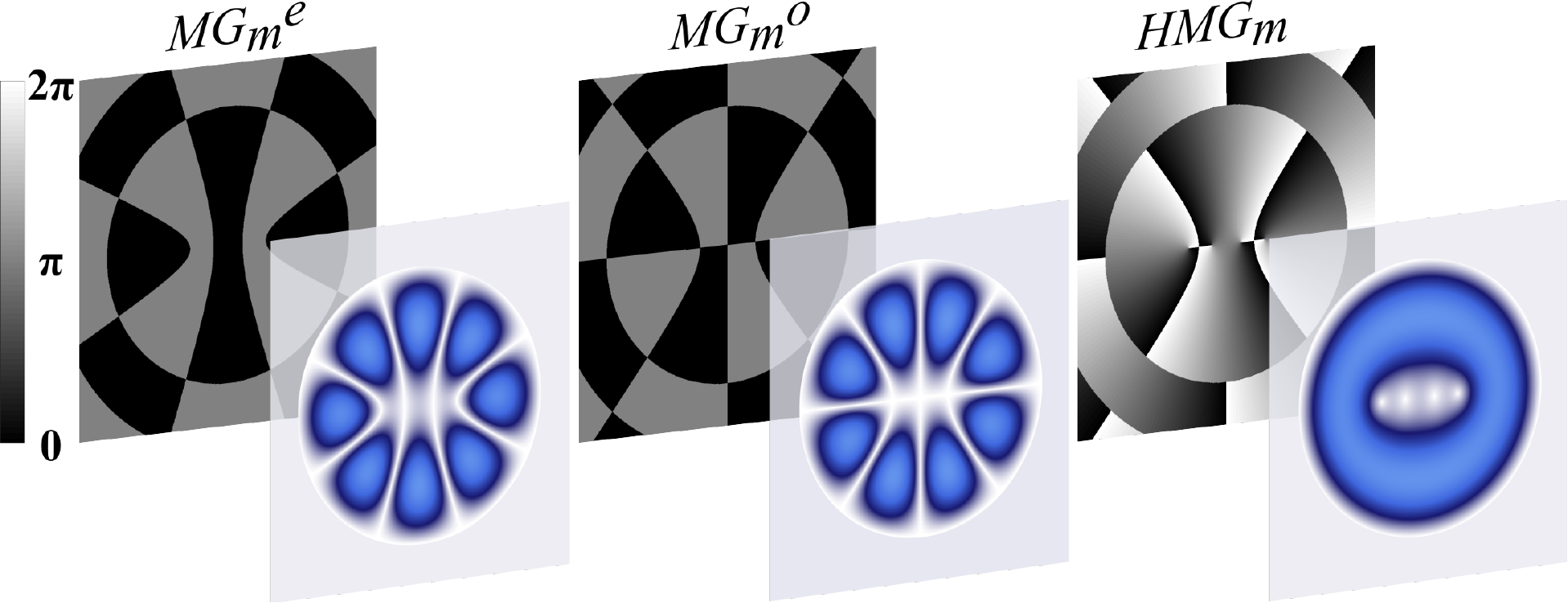}
    \caption{Transverse phase (back panels) and intensity profiles (front panels) of even (left), odd (middle) and helical (right) Mathieu-Gauss modes. For this example $a=1.5$, $k_t=4$, $e=0.9$ and $m=4$.}
    \label{ScalarMG}
\end{figure}

As an example, Fig.~\ref{ScalarMG} shows the intensity and phase profile of an even (left panel) and odd (middle panel) MG modes, respectively, while a HMG is shown on the right panel. Notice that the MG modes feature a petal-like intensity distribution, while the HMG shows a continuous intensity distribution. In a similar way, the MG modes are composed of discrete phase values $0$ and $\pi$, whereas the HMG modes carry a continuous phase distribution, with values in the interval $[0, 2\pi]$, that provides these with orbital angular momentum \cite{Chavez-Cerda2002}. 

\subsection{Helical Mathieu-Gauss vector beams}
\begin{figure}[b]
    \centering
    \includegraphics[width=0.49\textwidth]{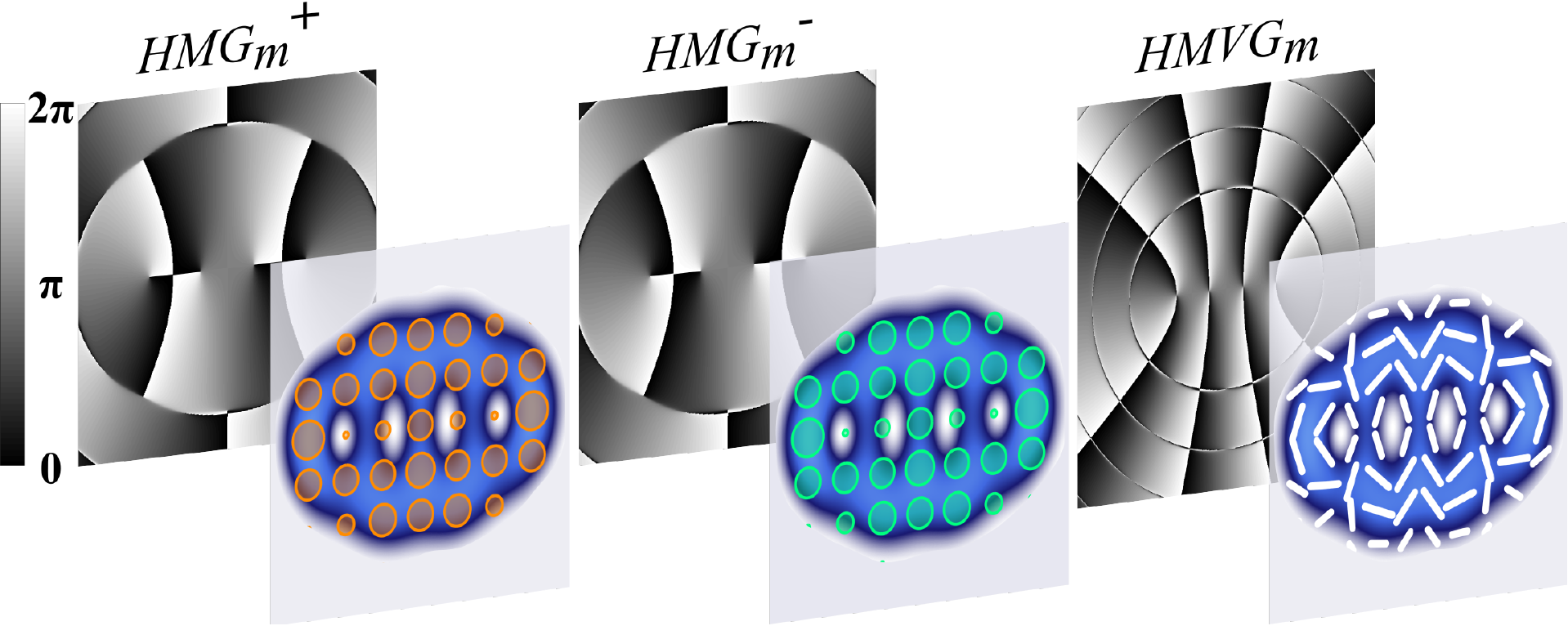}
    \caption{Transverse phase (back panels) and intensity profiles (front panels) of the scalar modes $\HMGp_4\hat{\bf e}_R$ and $\HMGm_4\hat{\bf e}_L$ that generate the vector mode, ${\bf HMGV}_{4,4}$, shown on the right. Right and left circular polarisation are represented by orange and green ellipses, respectively, while linear polarisation by white lines. For this example $a=1$, $k_t=3$, $e=0.9$ $m=4$ $\theta=\pi/4$ and $\alpha=0$.}
    \label{Vector modes}
\end{figure}

In this section we will introduce the Helical Mathieu-Gauss vector ($\HMGV$) beams, generated as a non-separable superposition of the polarisation and spatial degrees of freedom encoded in the previously described set of Helical Mathieu-Gauss beams. In general, vector beams are generated as a non-separable weighted superposition of the spatial and polarisation degrees of freedom as \cite{Galvez2012,Galvez2015}
\begin{equation}\label{Eq:VM}
    {\bf U}({\bf r})=\cos\theta u_R({\bf r})\hat{\bf e}_R+\sin\theta\exp(i\alpha) u_L({\bf r})\hat{\bf e}_L,
\end{equation}
where $\hat{\bf e}_R$ and $\hat{\bf e}_L$ are unitary vectors representing the right and left handed polarisation, respectively. The functions $u_R({\bf r})$ and $u_L({\bf r})$ represent the spatial degree of freedom, whose contribution is determined by $\theta\in[0,\pi/2]$. Finally, the exponential term $\exp(i\alpha)$, with $\alpha\in [0, \pi]$, is an inter-modal phase between both polarisation components. For the specific case of {\bf HMGV} modes, Eq.~\ref{Eq:VM} takes the form
\begin{equation}
    {\bf HMGV}_{m_1,m_2} = \cos\theta\: \HMGp_{m_1}\:\hat{\bf e}_R+\sin\theta\exp(i\alpha)\:\HMGm_{m_2}\:\hat{\bf e}_L,
    \label{HMVB}
\end{equation}
where from now on, we omit the explicit dependence on $(\xi,\eta,z;q)$, unless is necessary. The functions $\HMGp_{m}$ and $\HMGm_{m}$ represent the helical Mathieu-Gauss modes, given by Eq.~\ref{helical}, with right and left circular polarisation, respectively. To illustrate this conceptually, Fig.~\ref{Vector modes} shows the phase, polarisation and amplitude distribution of the scalar and vector modes represented by Eq.~\ref{HMVB}, for the specific case $a=1.5$, $k_t=4$, $e=0.9$,  $m=4$, $\theta=\pi/4$ and $\alpha=0$. Here, right and left circular polarisation states are represented by orange and green ellipses, respectively, while linear polarisation by white lines. It is worth mentioning that this polarisation representation corresponds to paraxial light beams. Front and back panels show, the intensity profile overlapped with the polarisation distribution and the phase of the mode, respectively, for the modes with right (left panel), $\HMGp_4$, and left (middle panel), $\HMGm_4$, circular polarisation, as well as their superposition (right panel).

\subsection{Representation of helical Mathieu-Gauss vector beams on the higher-order Poincar\'e sphere}
\begin{figure}[tb]
    \centering
    \includegraphics[width=0.49\textwidth]{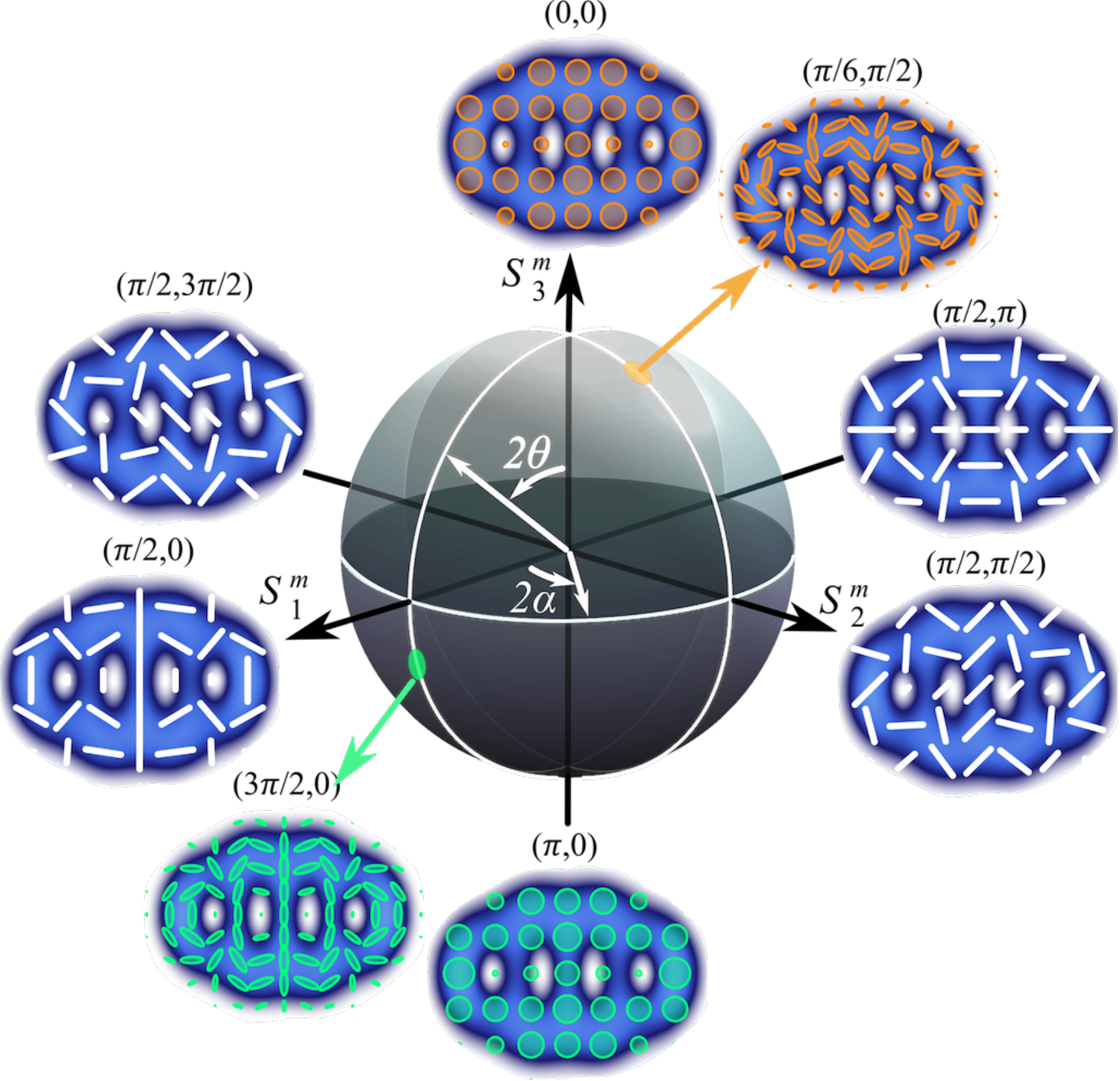}
    \caption{Geometric representation of ${\bf HMGV}_{m_1,m_2}$ modes on the higher-order Poincar\'e sphere. The scalar modes $\HMGp_m\hat{\bf e}_R$ and $\HMGm_m\hat{\bf e}_L$ are located in the North and South pole, respectively, while pure vector modes lie along the equator. Here we show the specific case for $a=1.5$, $k_t=4$, $e=0.9$  and $m=4$.}
    \label{MathieuSphere}
\end{figure}

It is well known that vector modes can be represented geometrically on a Higher-Order Poincar\'e Sphere (HOPS), in full analogy with the Poincar\'e representation of polarisation \cite{Milione2011}. In this representation, the scalar modes with right and left circular polarisation are mapped onto the North and South Poles, respectively, while pure vector modes are represented along the equator. In general, a given vector state with weight (a function of $\theta$) and inter-modal phase ($\alpha$) is represented as a point with coordinates $(2\theta,2\alpha)$. By way of example,  Fig.~\ref{MathieuSphere} shows the HOPS representation of the  mode ${\bf HMGV}_{4,4}$
with parameters $a=1.5$, $k_t=4$ and $e=0.9$. The insets illustrate the intensity and polarisation distribution of the scalar modes with specific coordinates $(0,0),(\pi,0),(\pi/2,0)$, $(\pi/2,\pi/2)$, $(\pi/2,\pi)$ and $(\pi/2,3\pi/2)$.
The Cartesian coordinates of this sphere are given by the higher-order Stokes parameters, $S_1^{m}$, $S_2^{m}$ and $S_3^{m}$, which for helical Mathieu vector modes are defined as
\begin{equation}
\begin{split}
    &S_0=|\HMGp_{m_1}|^2+|\HMGm_{m_2}|^2,\\
    &S_1=2|\HMGp_{m_1}||\HMGm_{m_2}|\cos\alpha,\\
    &S_2=2|\HMGp_{m_1}||\HMGm_{m_2}|\sin\alpha,\\
    &S_3=|\HMGp_{m_1}|^2-|\HMGm_{m_2}|^2,
    \end{split}
\end{equation}
where $S_0^2=S_1^2+S_2^2+S_3^2$. As we will show later, the Stokes parameters can be determined directly from a minimum of four intensity measurements \cite{Zhaobo2019} and can be used to reconstruct the transverse polarisation distribution as well as to determine the degree of coupling between the spatial and polarisation degrees of freedom \cite{Selyem2019,Manthalkar2020}. In addition, the parameters $\theta$ and $\alpha$ can also be directly determined from these parameters through the relations
\begin{equation}
\begin{split}
    &\alpha=\frac{1}{2}\arctan\left(\frac{S_2}{S_1}\right),\\
    &\theta=\frac{1}{2}\arcsin\left(\frac{S_3}{S_0}\right).
    \end{split}
    \label{Eq:Phase}
\end{equation}

\section{Experimental generation and characterisation of helical Mathieu-Gauss vector modes}
\label{Experiment}

\subsection{Experimental setup}
\begin{figure}[tb]
    \centering
    \includegraphics[width=0.49\textwidth]{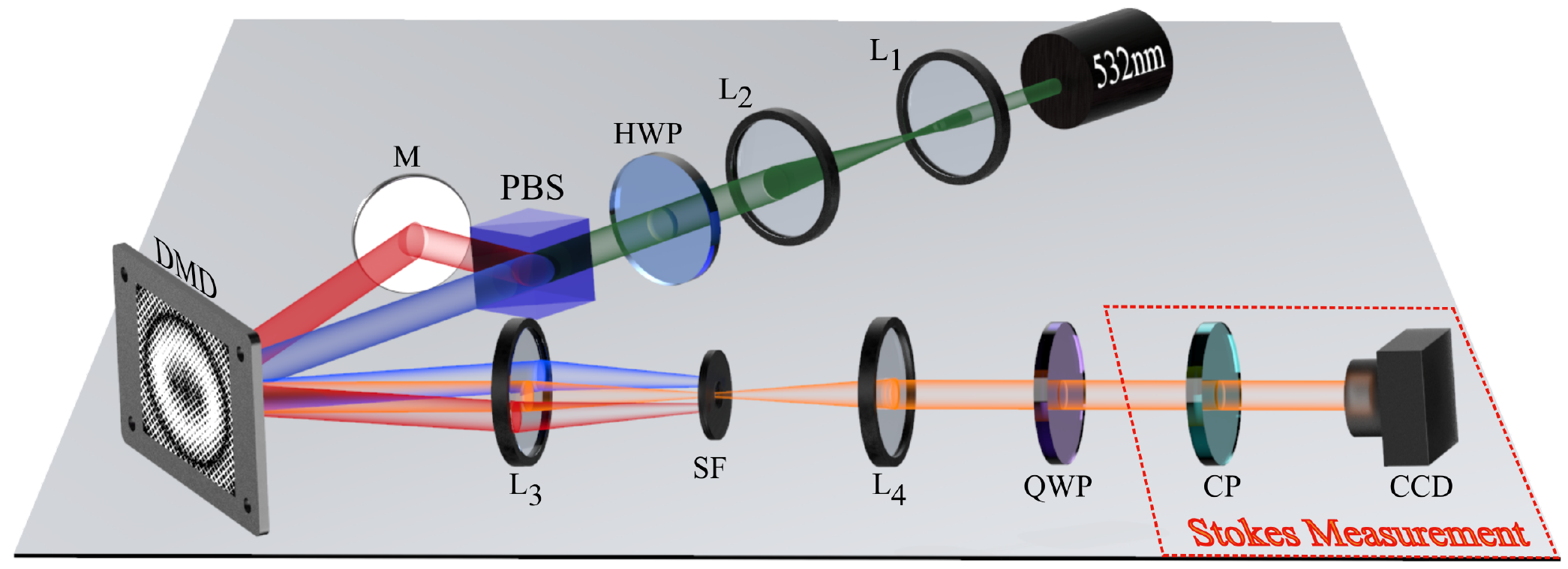}
    \caption{To generate the ${\bf HMGV}_{m_1,m_2}$ modes, a horizontally polarised beam, collimated and expanded by lenses L$_1$ and L$_2$, is transformed to diagonal polarisation with a half-wave plate (HWP) and split into its two polarisation components with a Polarising Beam Splitter (PBS). Both beams are redirected using a mirror (M) to the centre of a digital micromirror device (DMD), where they impinge at different angles. Two multiplexed holograms with unique linear phase gratings ensure the overlap on their first diffraction order along a common propagation axis, which is filtered with a Spatial Filter (SF) placed at the focusing plane of a telescope formed by lenses L$_3$ and L$_4$. A quarter-wave plate (QWP) transforms the beam from the linear to the circular polarisation basis. A Circular Polariser (CP) and a  charge-coupled device camera (CCD) measure intensities to compute the Stokes parameters.}
    \label{setup}
\end{figure}

The experimental setup implemented to generate arbitrary Mathieu vector modes is illustrated in Fig.~\ref{setup} and it comprises the use of a polarisation-insensitive Digital Micromirror Device (DMD). Here, a diagonally polarised laser beam ($\lambda=532$ nm) expanded and collimated by lenses L$_1$ and L$_2$ is split into its horizontal and vertical polarisation components via a Polarising Beam Splitter (PBS). Both beams are sent, one with the help of a mirror (M), to the centre of the DMD (DLP Light Crafter 6500 from Texas Instruments) impinging at slightly different angles ($\approx 1.5^\circ$). A multiplexed hologram containing the two independent holograms that generate each of the constituting wave fields of Eq.~\ref{HMVB} is displayed on the DMD. Each hologram is encoded with a unique linear phase grating, whose period is used to control the angle of propagation of each beam \cite{SPIEbook}. In this way, the first diffraction order of each beam propagates along a common axis, where the vector beam is generated (see \cite{Rosales2020} for more details). Higher diffraction orders are removed with a telescope ($\rm L_3$ and $\rm L_4$) in combination with a spatial filter (SF) located at the focusing point of the telescope. In addition, a Quarter-Wave plate (QWP$_1$) oriented at $45^\circ$ is placed afterwards to change the vector beam from the linear to the circular polarisation basis. Finally, the beam is sent to a Charge-Coupled Device camera (CCD, $1.55\mu$m pixel size), which in combination with a Circular Polariser (CP) allows us to measure the required intensities to compute the Stokes parameters, as we will explain in the following section. 

\subsection{Stokes parameters, polarisation and concurrence}\label{characterisation}
It is well-known that Stokes polarimetry allows to reconstruct the transverse polarisation distribution of a given mode. In addition, it also allows to determine the degree of coupling between the scalar and polarisation degrees of freedom using a basis-independent tomography \cite{Selyem2019}. As such, in this section we will briefly describe how the Stokes parameters can be determined experimentally. To start with, let us remind that these parameters can be determined from four intensities measurements as \cite{Goldstein2011},
\begin{equation}\label{Eq:Stokes}
\begin{split}
\centering
     &S_{0}=I_{H}+I_{V},\\
     &S_{1}=2I_{H}-S_{0},\\
     &S_{2}=2I_{D}-S_{0},\\
     &S_{3}=2I_{R}-S_{0},
\end{split}
\end{equation}
\begin{figure}[t]
    \centering
    \includegraphics[width=0.49\textwidth]{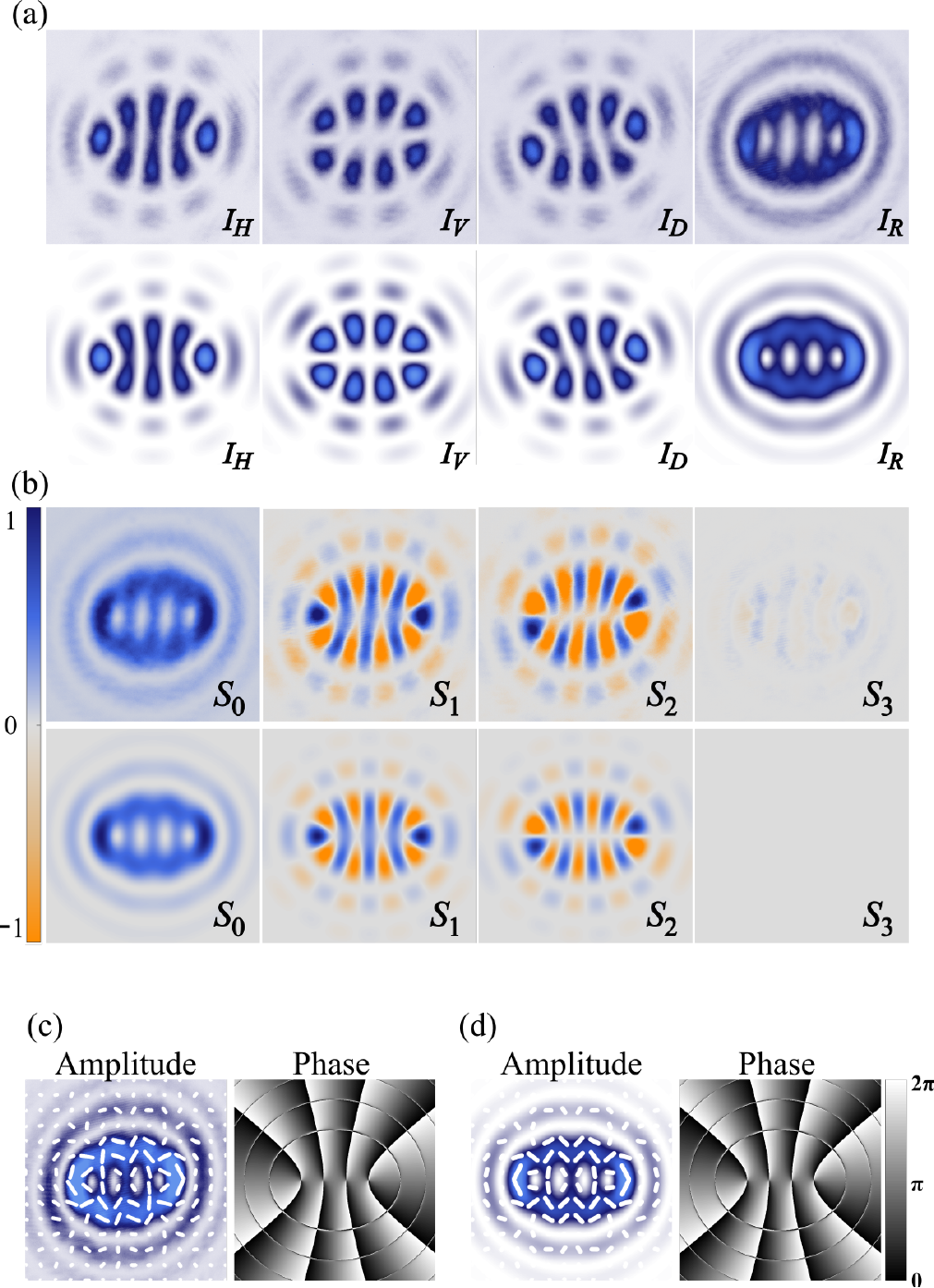}
    \caption{(a) Intensity distribution of the polarisation components $I_H$, $I_V$, $I_D$ and $I_R$, experiment on top and theory on the bottom. (b) Stokes parameters computer from these intensities, experiment on top and theory on the bottom. Experimental (c) and theoretical (d) reconstructed polarisation distribution along the transverse plane  (left panel) along with the reconstructed phase of the complex Stokes field (right panel). This example corresponds to the mode ${\bf HMGV}_{m_1,m_2}(\xi,\eta,z,q)$ with parameters $a=1.5$, $k_t=4$, $e=0.9$, $m_1=m_2=4$, $\theta=\pi/4$ and $\alpha=0$.}
    \label{Fig:Intensity}
\end{figure}\\
where $I_H$, $I_V$, $I_D$ and $I_R$ represent the intensity of the horizontal, vertical, diagonal and right-handed polarisation components, respectively. These intensities can be measured using a linear polariser and a set of phase retarders but, since they can introduce intensity mismatches, we favoured the use of a Circular Polariser (CP). Such CP is made by cementing a QWP to a linear polariser with its axis at $45^{\circ}$ with respect to the fast axis of the QWP. Using this optical component, $I_H$, $I_V$ and $I_D$ can be obtained by setting the CP to $0^{\circ}$, $90^{\circ}$ and $45^{\circ}$, respectively. Finally, $I_R$ is obtained by flipping the CP $180^{\circ}$ while setting its angle to  $0^{\circ}$. An alternative way to measure all intensities simultaneously, while also avoiding intensity mismatches was reported recently and is based on the use of a DMD \cite{Zhaobo2019}. Figure \ref{Fig:Intensity}(a) shows an example of the experimental (top) and theoretical (bottom) expected intensities for the specific case ${\bf HMGV}_{m_1,m_2}(\xi,\eta,z=0;q)$, with parameters $a=1.5$, $k_t=4$, $e=0.9$, $m_1=m_2=4$, $\theta=\pi/4$ and $\alpha=0$. The Stokes parameters can then be computed from Eq.~\ref{Eq:Stokes} using these intensity images, which are displayed in Fig \ref{Fig:Intensity}(b), experiment on top and theory on the bottom. 


\subsection{Polarisation reconstruction}
Once the Stokes parameters have been determined, it is straight forward to reconstruct the entire transverse polarisation distribution. For this, it is worth reminding that the parameters defining a polarisation ellipse can be directly computed from the Stokes parameters (see for example \cite{Goldstein2011}). Now, polarisation ellipses are reconstructed on a $N\times N$ grid ($N=14$ in our case) of equally spaced points across the transverse plane taken from the experimental results. An example of the experimental polarisation distribution reconstructed from the Stokes parameters of Fig.~\ref{Fig:Intensity}(b) is shown in the left panel of Fig.~\ref{Fig:Intensity}(c). For comparison, the left panel of Fig.~\ref{Fig:Intensity}(d) shows the reconstructed polarisation from the theoretical Stokes parameters. Figure \ref{Fig:Intensity} also shows the experimental and theoretical phase of the complex Stokes field, reconstructed from the argument of $S = S_{1} + iS_{2}$.

\subsection{Degree of concurrence}\label{sec:Cdef}
The polarisation distribution provides with a qualitative measure of whether a given mode is scalar or vector. However, it does not provide any information about its degree of "vectorness", {\textit i.e.}, degree of coupling between both degrees of freedom. Crucially, and given the similarity of vector modes with local quantum entanglement, this can be determined through the concurrence ($C$) \cite{McLaren2015,Ndagano2016,Yao-Li2020,Selyem2019}. The concurrence takes values in the interval $[0,1]$, assigning 0 to scalar modes and 1 to vector modes.  $C$ can be determined from the stokes parameters using the relation \cite{Selyem2019,Manthalkar2020},
\begin{equation}
C=\sqrt{1-\left(\frac{\mathbb{S}_1}{\mathbb{S}_0} \right)^2-\left(\frac{\mathbb{S}_2}{\mathbb{S}_0} \right)^2-\left(\frac{\mathbb{S}_3}{\mathbb{S}_0} \right)^2},
\label{concurrence}
\end{equation}
where $\mathbb{S}_i$ are the values of the Stokes parameters $S_i$ integrated over the entire transverse profile, that is, 
\begin{equation}
\mathbb{S}_i=\iint_{-\infty}^\infty S_{i} dA \qquad i=0,1,2,3
\end{equation}
By way of example, integration over the entire plane of the Stokes parameters shown in Fig.~\ref{Fig:Intensity} yields the values $\mathbb{S}_1/\mathbb{S}_0=0.04$, $\mathbb{S}_2/\mathbb{S}_0=0.08$ and $\mathbb{S}_3/\mathbb{S}_0=-0.02$, which after substitution into Eq.~\ref{concurrence} gives $C=0.99$. The methods described in this section will be used in the next section, where we will present our main findings.

\section{Experimental results}\label{results}

\subsection{Generation of helical Mathieu-Gauss vector beams on the HOPS}
\begin{figure}[tb]
    \centering
    \includegraphics[width=0.49\textwidth]{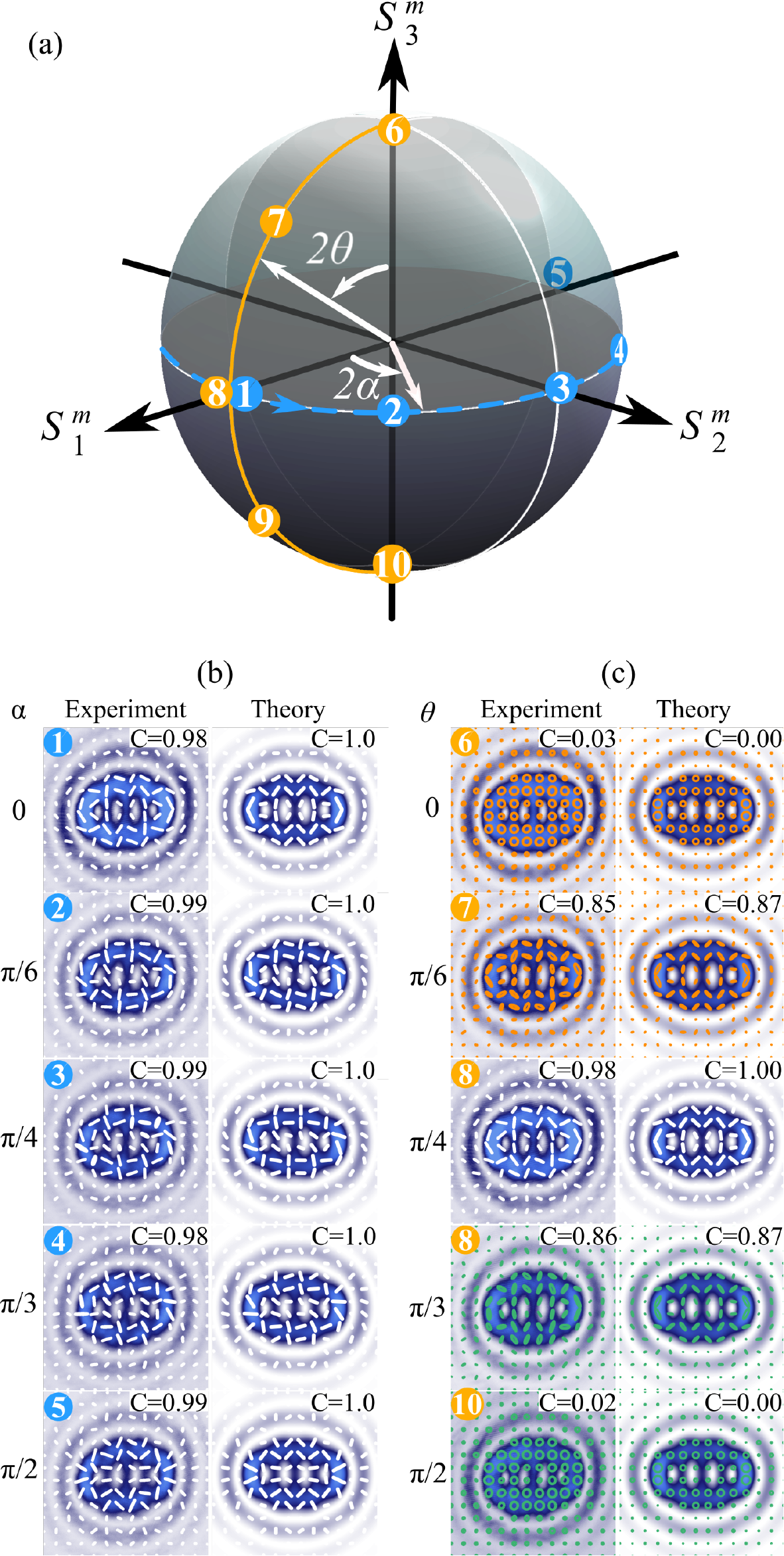}
    \caption{(a) Geometric representation of ${\bf HMGV}_{m,m}(\xi,\eta,z=0,q)$ vector modes on the HOPS. The modes with varying degrees  of concurrence are represented along the solid yellow line and given by the specific values $2\theta=0, \pi/6, \pi/2, \pi/3, \pi$ and $\alpha=0$. The modes with varying inter-modal phase are represented along the dashed blue line along the equator, for which, $2\alpha=0, \pi/3, \pi/2, 2\pi/3, \pi$ and $\theta=\pi/4$. Intensity  and  polarisation  distribution  of  modes  with, different inter-modal phases (b) and different  degrees  of  non-separability (c). For each case, the theoretical predictions are shown on the right while the experimental results on the left. The numbers inserted in the top-left corner indicate their positions on the HOPS.}
    \label{HOPS}
\end{figure}

We start the description of our results with experimentally generated ${\bf HMGV}$ modes on the higher-order Poincar\'e sphere. Without the loss of generality, we will restrict this analysis to the specific case ${\bf HMGV}_{m,m}(\xi,\eta,z=0;q)$, with parameters $a=1$, $k_t=6$, $e=0.9$ and $m=4$. Further, we will only show a set of ten representative modes, five along the equator and five more along a path connecting the poles of the HOPS, as shown is Figure \ref{HOPS} (a). The modes along the equator, which correspond to pure vector modes, are represented with numbers from 1 to 5 and connected with a blue dashed line. Their specific coordinates are $(\pi/2,0)$, $(\pi/2,\pi/3)$, $(\pi/2,\pi/2)$, $(\pi/2,2\pi/3)$ and $(\pi/2,\pi)$, respectively. Similarly, the set of modes along the meridian, with specific coordinates $(\pi/2,0)$, $(\pi/2,\pi/6)$, $(\pi/2,\pi/2)$, $(\pi/2,\pi/3)$ and $(\pi/2,\pi)$, are represented with numbers from 6 to 10 and connected with a solid yellow line.
The transverse intensity profile overlapped with the corresponding polarisation distribution of the set pure vector modes (lying along the equator of the HOPS) are shown in Fig.~\ref{HOPS}(b), experiment on the left and theory on the right. For all these modes $\theta=\pi/4$, $\alpha=0,\pi/6,\pi/4,\pi/3,\pi/2$, as shown on the left-hand side of each panel and labelled with numbers from 1 to 5. As can be seen, an increment in the inter-modal phase has the only effect of rotating the polarisation distribution. Notice the high similarity between the experimental and theoretical intensity and polarisation distributions. The corresponding concurrence, computed as described in section \ref{sec:Cdef}, is also included in the top-right corner of each mode. Notice that $C$ is higher than $0.98$ for all modes, which implies our generated vector modes have a high purity. In a similar way, the case of modes with different degrees of non-separability and fixed inter-modal phase ($\alpha=0$), represented along the path connecting the North and South Poles of the HOPS, are displayed in Fig.~\ref{HOPS}(c), experiment on the left and theory on the right. These modes are also labelled with numbers from 6 to 10 (top-left corner inset) just as is the HOPS in Fig.~\ref{HOPS}(a). Their calculated concurrence values are also displayed as an inset on the top-right corner of all modes. Notice the transition from the right-handed ($\theta=0$) to the left-handed homogeneously polarised mode ($\theta=\pi/2$), passing through a pure vector mode at $\theta=\pi/4$.

\subsection{Helical Mathieu-Gauss vector modes of different eccentricity}
\begin{figure}[tb]
    \centering
    \includegraphics[width=0.49\textwidth]{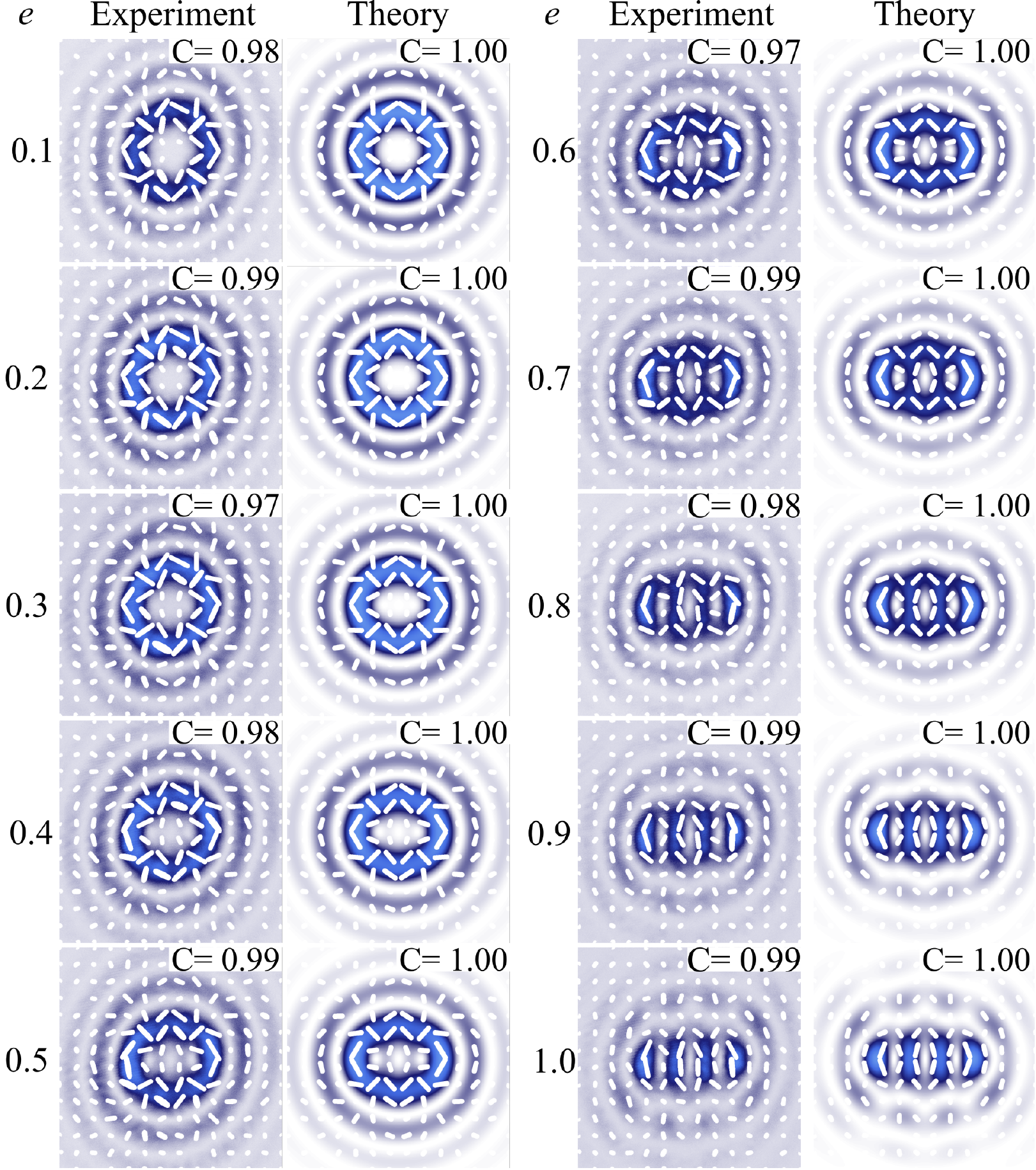}
    \caption{ Polarisation distribution of helical Mathieu vector modes as function of the ellipticity $\varepsilon$.   Experimental (left) and theoretical (right) transverse intensity profile overlapped with the polarisation distribution. Here we show the specific case $m=3$, $k_t=6$ and $a=1$ for $e\in[0.1-1.0]$.}
    \label{Ecxentricity}
\end{figure}
In this section we analyse ${\bf HMGV}$ modes with increasing values of eccentricity $e$. Remember that the ellipticity of the mode increases in proportion to $e$, from circular ($e=0$) to elliptical ($e>0$), which, as we will show, also affects the polarisation distribution. We will restrict our description to the representative set of pure vector modes ${\bf HMGV}_{m,m}(\xi,\eta,z=0,q)$ with parameters $m=3$, $k_t=6$, $a=1$, $\alpha=0$ and $\theta=\pi/4$. Figure \ref{Ecxentricity} shows our experimental results for eccentricity values $e \in[0.1,1.0]$, where we show the intensity profile overlapped with the corresponding reconstructed polarisation distribution. The first two columns show the modes with eccentricities $e\in[0.1,0.5]$, while the last two columns show the modes with eccentricities $\varepsilon\in[0.6,1.0]$. In both cases the experimental results are on the left panel and the theoretical on the right one. The calculated concurrence values $C$ are shown as insets in the top-right corner of each panel. Notice that the ${\bf HMGV}$ modes maintain a maximum value of concurrence for all eccentricity values, $c=1$ for the theoretical cases and higher than 0.97 for the experimental cases.

\subsection{Helical Mathieu-Gauss vector beams of higher orders}
\begin{figure}[b]
    \centering
    \includegraphics[width=0.49\textwidth]{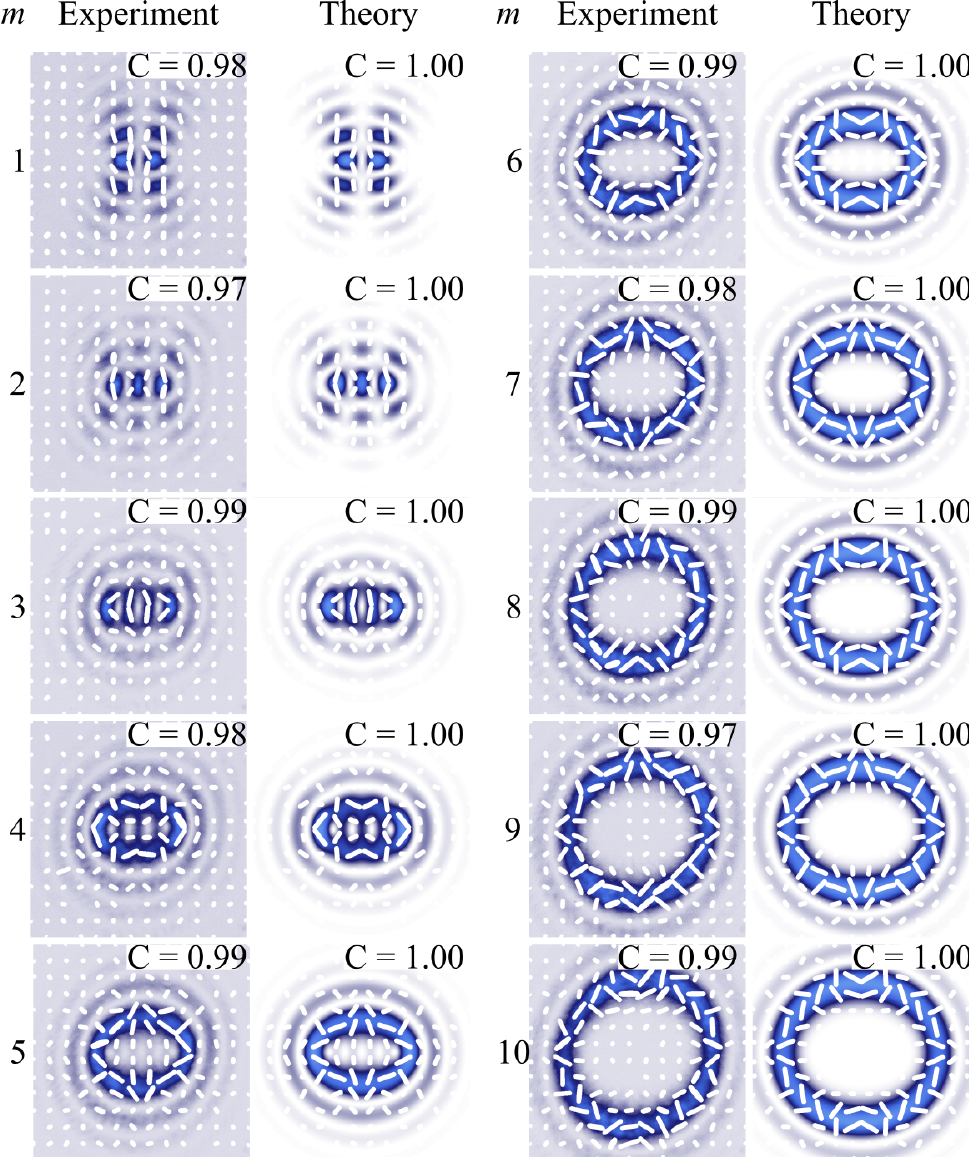}
    \caption{Polarisation distribution of helical Mathieu-Gauss vector modes ${\bf HMV}_m(\xi,\eta,z=0,q)$ with parameters $k_t=6$, $a=1$, $e=0.9$ and $\alpha=0$, as function of the order $m$.}
    \label{OAM}
\end{figure}

In this section we present our results concerning the helical Mathieu-Gauss vector modes of higher orders, defined by the parameter $m$. A representative set of pure $\HMGV_{m,m}$ modes with parameters $k_t=6$, $a=1$, $e=0.9$, $\alpha=0$ and $\theta=\pi/4$ is shown in Fig.~\ref{OAM} for $m\in[1,10]$. The first two columns show the modes with values $m\in[1,5]$, experiment on the left and theory on the right, while the last two columns show the modes for  $m\in[6,10]$. Each panel shows the intensity profile and overlapped transverse polarisation distribution. For this specific parameters set, the modes evolve from a vertical structure of multiple intensity lobes ($m\leq4$) to a single horizontal elliptical structure ($m>4$), which increases in size with $m$. The polarisation distribution also evolves from nearly vertical polarisation ($m=1$) into a more intricate structure at $m=10$. As in the previous cases, here $C=1$ for all the theoretical modes and higher that 0.97 for the experimental ones, as shown in the top-right corner of each panel.
\subsection{Propagation of Helical Mathieu-Gauss vector modes}
\begin{figure}[tb]
    \centering
    \includegraphics[width=0.49\textwidth]{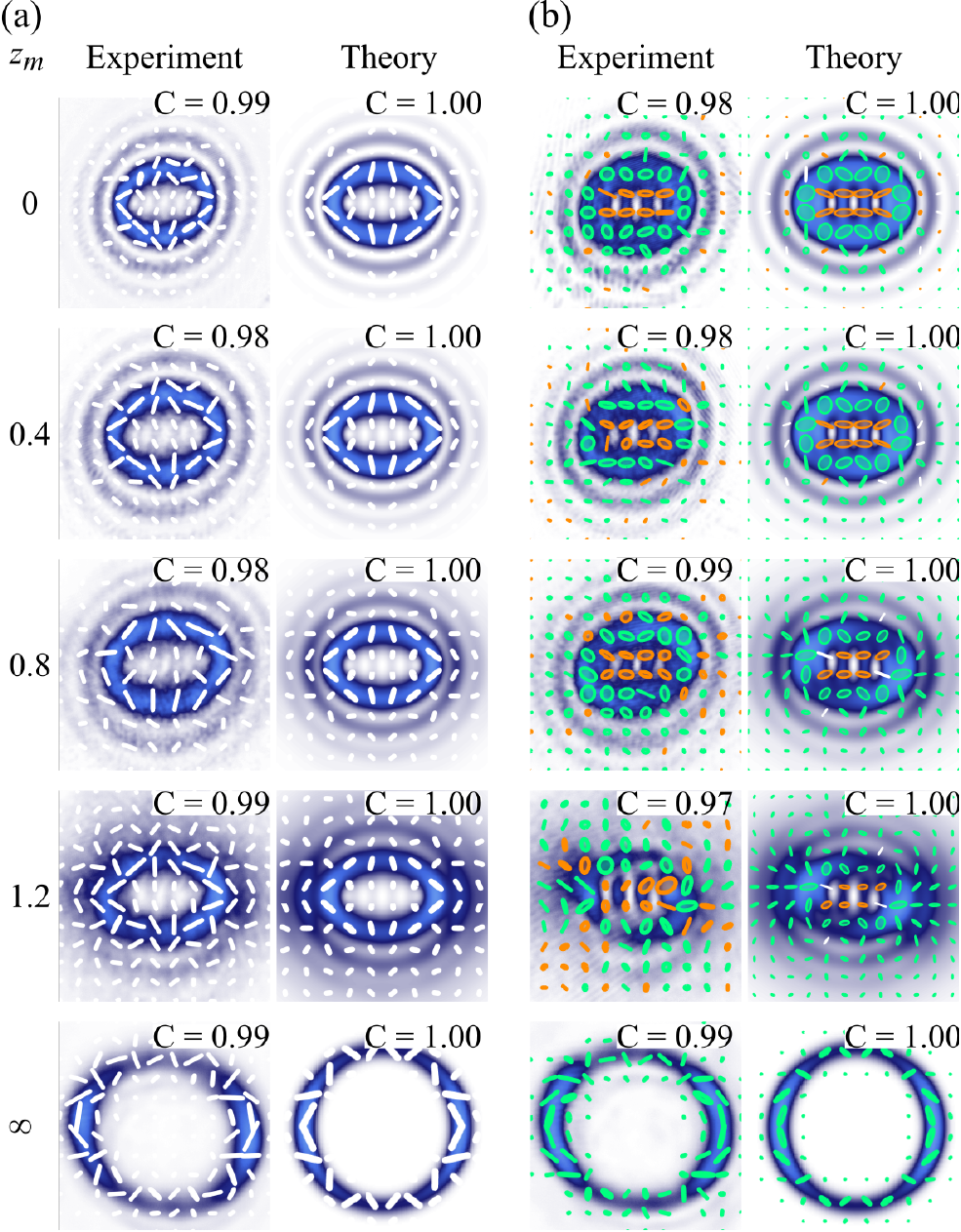}
    \caption{Helical Mathieu-Gauss vector modes as function of propagation distance ${\bf HMGV}_m(\xi,\eta,z,q)$ with parameters $k_t=6$, $a=1$, $e=0.9$, $\theta=\pi/4$ and $\alpha=0$, as function of the propagation distance $z$. In (a) $m_1=m_2=5$ and in (b) $m_1=3$, $m_2=5$.}
    \label{Prop}
\end{figure}

As a final case, in this section we present an analysis of the ${\bf HMGV}_{m_1,m_2}$ modes as function of the propagation distance $z$. Without the loss of generality, we restrict our analysis to the set of modes defined by the parameters $k_t=6$, $a=1$, $e=0.9$, $\theta=\pi/4$ and $\alpha=0$. We first perform a qualitative analysis based on the reconstruction of their polarisation distribution at the  planes $z=0$, $z=0.4 z_m$, $z=0.8z_m$, $z=1.2z_m$ and $z=\infty$ (far field). Followed by a quantitative analysis through the concurrence $C$, computed at the same planes, to determine if the purity of the mode changes over propagation. In Fig.~\ref{Prop} (a) we show the polarisation distribution overlapped on the corresponding intensity profile for the case $m_1=m_2=5$ with experiment on the left and theory on the right. As expected, the intensity distribution remains almost invariant for $z<z_m$, whereas for $z>z_m$ diffraction effects become dominant. Finally, in the far-field it acquires a ring-shaped intensity profile. In an analogous way, the polarisation distribution also remains invariant in the non-diffracting region, maintaining a similar distribution in the far field. The concurrence remains higher that $C=0.98$ for all propagation planes, an indication that the purity of the vector mode remains invariant upon propagation. Finally, in Fig.~\ref{Prop} (b) we show the case of modes with different order, $m_1=3$ and $m_2=5$, experiment on the left and theory on the right. Here, even though the intensity distribution remains almost the same in the non-diffracting region, the polarisation distribution experiences noticeable changes, which we expect will increase as the difference $|m_2-m_1|$ increases. Crucially, the concurrence remains unmodified, higher than $0.98$ for all planes, indicating that even though the polarisation distribution changes, the degrees of freedom remain maximally coupled.

\section{Conclusions}
In summary, in this manuscript we introduce, theoretically and experimentally, the helical Mathieu-Gauss vector modes with an elliptical spatial distribution, generated as a weighted superposition of orthogonal helical Mathieu-Gauss beams. As we show, these modes can be represented geometrically in the Higher-Order Poincar\'e Sphere (HOPS) with the scalar modes located on the sphere's poles and pure vector modes along the equator. Furthermore, we demonstrated their experimental generation using a compact experimental setup utilising a DMD (although an SLM can also be used). Such generation technique allows us to generate helical Mathieu vector modes with arbitrary polarisation distribution and degree of concurrence. We also generated other sets of helical Mathieu vector modes with different values of eccentricity, which we varied from 0 to 1. Finally, we analysed the case of vector modes of different orders, $m\in[1,10]$. We analysed qualitatively and quantitatively the generated modes by reconstructing their transverse polarisation distribution and measuring the non-separability between the spatial and polarisation degrees of freedom through the concurrence $C$ . Finally, it is worth mentioning that scalar Mathieu beams have already demonstrated their potential in applications such as a optical tweezers and optical communications. It is therefore likely that the helical Mathieu vector modes presented here will pioneer novel applications that will not be limited to these fields. Our demonstration paves the path towards the use of vector beams with elliptical symmetries in research fields such as laser material processing, optical communications, imaging, optical manipulations, among others.

\section*{Funding}
 This work was partially supported by the National Nature Science Foundation of China (NSFC) under Grant No.  61975047.  BPG and RIHA acknowledge support from Consejo Nacional de Ciencia y Tecnología (PN2016-3140).
\section*{Disclosures}
The authors declare that there are no conflicts of interest related to this article.
\section*{References}
\bibliographystyle{iopart-num}
\providecommand{\newblock}{}

\end{document}